# Josephson effect in a weak link between borocarbides

Yu.A. Kolesnichenko and S.N. Shevchenko

*B. Verkin Institute for Low Temperature Physics and Engineering*
*of the National Academy of Sciences of Ukraine, 47 Lenin Ave., Kharkov 61103, Ukraine*
E-mail: kolesnichenko@ilt.kharkov.ua

A stationary Josephson effect is analyzed theoretically for a weak link between borocarbide superconductors. It is shown that different models of the order parameter result in qualitatively different current-phase relations.

PACS: **74.50.+r**

Determination of the symmetry of the order parameter $\Delta$ in novel unconventional superconductors is important for the development of modern physics of superconductivity because the dependence of $\Delta(\mathbf{k})$ on the direction of the electron wave vector $\mathbf{k}$ on the Fermi surface determines all of the kinetic and thermodynamic characteristics of the superconductor. Calculation of the order parameter $\Delta(\mathbf{k})$ is a fundamental problem and requires knowledge of the pairing potential. Some general information about $\Delta(\mathbf{k})$ can be obtained from the symmetry of a normal state, i.e., according to the Landau theory of secon-order phase transitions [1], the order parameter transforms only accoding to irreducible representations of the symmetry group of the normal state (for review, see [2]). Nevertheless, symmetry considerations reserve for the order parameter considerable freedom in the selection of irreducible representation and its basis functions. Therefore in many papers authors consider different models of the order parameter, which are based on possible representations of crystallographic point groups. The subsequent comparison of theoretical results with experimental data makes it possible to choose between available models of the order parameter. The Josephson effect in superconducting weak links is one of the most suitable instruments for investigation of the symmetry of $\Delta(\mathbf{k})$. It has heen shown, for example, that current–phase relations $j_J(\varphi)$ in unconventional superconductors are quite different for different models of $\Delta(\mathbf{k})$, and hence the study of the Josephson effect enables one to judge the applicability of different models to the novel superconductors [3].

Borocarbides, such as $YNi_2B_2C$ and $LuNi_2B_2C$, exhibit unconventional superconductivity. There is strong evidence that in these materials the order parameter is highly anisotropic [4]. The order parameter in these compounds has fourfold symmetry, and there are deep minima along the [100] and [010] directions [4,5]. Both the symmetry of the borocarbide crystal structure and the experimental results have allowed the authors of Refs. [6] to suggest an $s + g$-wave model of the order parameter to describe the superconductivity in the borocarbides:

$$\Delta = \Delta_s - \Delta_g \sin^4 \vartheta \cos 4\varphi \equiv \frac{\Delta_0}{2}(\gamma - \sin^4 \vartheta \cos 4\varphi), \quad (1)$$

where $\vartheta$ and $\varphi$ are the polar and azimuthal angles of an electron wave vector $\mathbf{k}$; $\Delta_s$ and $\Delta_g$ are the $s$ and $g$ components of the order parameter, and $\Delta_0 = \Delta_0(T)$ describes the temperature-dependent amplitude value of the order parameter.

Parameter $\gamma = \Delta_s / \Delta_g$ is the key value here. If $\gamma < 1$, then the order parameter $\Delta(\vartheta, \varphi)$ is an alternating-sign quantity, which means that some reflected trajectories experience the intrinsic phase difference. This result in the suppression of the order parameter in the vicinity of the interface between two superconductors similar to what is known about the contact of two $d$-wave superconductors (see [7] and references therein); and in this case the non-self-consistent calculation, presented below, can be justified for the weak links in the form of both the point contact and the plane boundary between two banks. Another consequence of the intrinsic phase difference is the appearance of the spontaneous phase difference (which means that at equilibrium, when $j_J = 0$ and $dj_J / d\phi > 0$, the phase difference is not zero: $\phi = \phi_0 \neq 0$) and the spontaneous interface current at equilibrium at $\phi = \phi_0$ (which is demonstrated below). If $\gamma \geq 1$, then the order parameter is not an alternating-sign quantity, $\Delta(\vartheta, \varphi) \geq 0$, and the non-self-consistent calculation can only be justified for the weak link in the form of the point contact. In this case at the contact there is also the component of the current along the interface due to the anisotropy of the order parameter. However this current is not spontaneous, which means that at the equilibrium at $\phi = 0$ both Josephson and interface current components equal to zero.

In what follows we study the stationary Josephson effect in the weak link between two borocarbides, described by the $s + g$-wave model (1) of the order parameter, and compare the results with the Josephson current between $d$-wave superconductors ($\Delta = \Delta_0 \sin 2\varphi$). We consider a perfect contact between two clean, differently orientated superconductors. The external order parameter phase difference $\phi$ is assumed to drop at the interface plane $x = 0$. The theoretical description of the Josephson effect is based on the Eilenberger equation, as it was described, for example, in the Refs [7]. The standard procedure of matching the solutions of the bulk Eilenberger equations at the boundary gives the Matsubara Green's function $\hat{G}_\omega(0)$ at the contact at $x = 0$ [7]. The component $G_\omega^{11}(0) \equiv g_\omega(0)$ of $\hat{G}_\omega(0)$ determines the current density at the boundary:

$$\mathbf{j}|_{x=0} = -j_\Delta 4\pi \sum_{\omega>0} \langle \hat{\mathbf{k}} \operatorname{Im} g(0) \rangle_{\hat{\mathbf{k}}}, \quad j_\Delta = |e| N_0 v_F \Delta_0 \quad (2)$$

$$\operatorname{Im} g(0) = -\operatorname{sign}(k_x) \frac{\Delta_L \Delta_R \sin\phi}{\Omega_L \Omega_R + \omega_n^2 + \Delta_L \Delta_R \cos\phi}, \quad (3)$$

where $N_0$ is the density of states at the Fermi level, $\langle \ldots \rangle_{\hat{\mathbf{k}}}$ denotes averaging over the directions of Fermi wave vector $\mathbf{k}$, $\hat{\mathbf{k}} = \mathbf{k}/k$ is the unit vector in the direction of $\mathbf{k}$, $\omega_n = \pi T(2n+1)$ are Matsubara frequencies, $\Delta_{L,R}$ stands for the order parameter in the left (right) bank, and $\Omega_{L,R} = \sqrt{\omega_n^2 + \Delta_{L,R}^2}$.

Making use of Eqs. (2), (3) we numerically plot the current–phase relations for two components of the current, $j_x$ (through the contact) and $j_y$ (along the contact). We assume that the $c$ axes of the left and right superconducting banks are directed along the $z$ axis, that the $a$ and $b$ axes of the left superconductor are directed along $x$ and $y$ axes, and that the $ab$ basal plane of the right superconductor is rotated by an angle $\alpha$ with respect to the left superconductor. In Fig. 1 the current through the contact (Josephson current) is plotted versus the phase difference for both $d$-wave and $s + g$-wave models of the order parameter for low temperature and a relative angle between superconducting banks $\alpha = \pi/4$. The current–phase relations are *qualitatively* different, which can be used as a test to discriminate experimentally the true model, that describes a borocarbide.

Generally speaking, there is a current $j_y$ tangential to the boundary in addition to the current through the contact $j_x$ [3]. The case of the contact of two $d$-wave superconductors has been considered in many papers (see [7] and references therein). Here we consider in more detail the Josephson effect for the $s + g$-wave model. In this case the Josephson current depends weakly on the relative angle between superconductors $\alpha$, while the current along the contact plane depends strongly on $\alpha$: $j_y = 0$ for $\alpha = 0$ and $\pi/4$ and attains the maximal value at $\alpha = \pi/8$. In Fig. 2 we plot both $j_x$ and $j_y$ for the $s + g$-wave model of the order parameter for low temperature and $\alpha = \pi/8$. If $\gamma \geq 1$, then the tangential component of the current density in the contact plane remains much smaller than the transverse component for any values of the phase difference $\phi$ and of the relative angle between superconductors $\alpha$. At $\gamma < 1$, as it is pointed out above, the spontaneous phase difference and spontaneous interface current appear. The effect is the most pronounced at $\gamma \ll 1$, which we illustrate at Fig. 2.

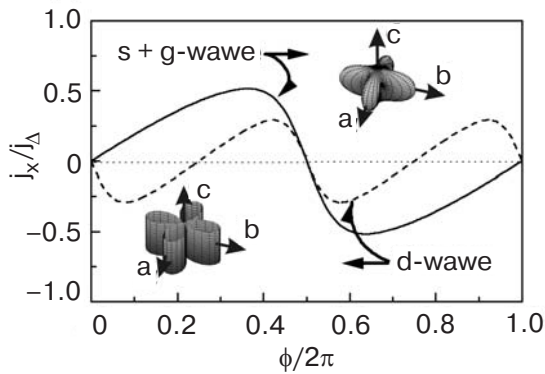

*Fig. 1.* Josephson current density versus phase difference for both the $d$-wave and $s + g$-wave models of the order parameter (the solid line corresponds to $\gamma = 1$ and the dotted line corresponds to $\gamma = 2$). $T = 0.05\Delta_0$, $\alpha = \pi/4$. The order parameters for the $d$-wave and $s + g$-wave models in momentum space are shown in the insets.

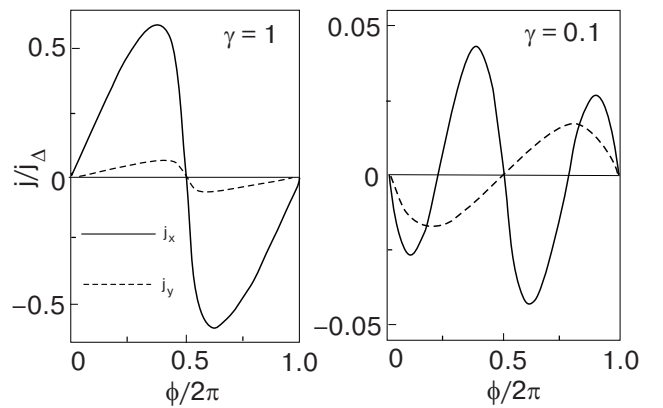

*Fig. 2.* Current–phase relations for two components of the current, $j_x$ (through the contact) and $j_y$ (along the contact) for $\gamma = 1$ and $\gamma = 0.1$, $T = 0.05\Delta_0$, $\alpha = \pi/8$.

Thus we have considered a weak link between two clean differently orientated borocarbide superconductors. The current–phase relations were compared for the $d$-wave and $s + g$-wave models of the order parameter. The dependences of the Josephson current on the phase difference are qualitatively different for these models. It is shown that because of the anisotropy of the order parameter there is a current tangential to the boundary for the $s + g$-wave model, which attains its maximum at a relative angle between superconductors equal to $\pi/8$. This interface current can exist in the absence of Josephson current at equilibrium if $\gamma < 1$. The observation of such spontaneous current can be used as a test of whether the order parameter is alternating-sign or not.

We acknowledge fruitful discussions with A.N. Omelyanchouk. This work was supported by CRDF Project (Grant No UP1-2566-KH-03).

1. L.D. Landau and E.M. Lifshitz, *Statistical Physics*, Part 1, Pergamon, New York (1979).
2. V.P. Mineev and K.V. Samokhin, *Introduction to Unconventional Superconductivity*, Amsterdam, The Netherlands: Gordon and Breach science Publishers (1999).
3. Yu.A. Kolesnichenko, A.N. Omelyanchouk, and A.M. Zagoskin, *Fiz. Nizk. Temp.* **30**, 714 (2004) [*Low Temp. Phys* **30**, 535 (2004)].
4. Etienne Boaknin, R.W. Hill, Cyril Proust, C. Lupien, Louis Taillefer, and P.C. Canfield, *Phys. Rev. Lett.* **87**, 237001 (2001); K. Izawa, K. Kamata, Y. Nakajima, Y. Matsuda, T. Watanabe, M. Nohara, H. Takagi, P. Thalmeier, and K. Maki, *Phys. Rev. Lett.* **89**, 137006 (2002).
5. T. Jacobs, Balam A. Willemsen, S. Sridhar, R. Nagarajan, L.C. Gupta, Z. Hossain, Chandan Mazumdar, P.C. Canfield, and B.K. Cho, *Phys. Rev.* **B52**, 007022 (1995); In-Sang Yang, M.V. Klein, S.L. Cooper, P.C. Canfield, B.K. Cho, and Sung-Ik Lee, *Phys. Rev.* **B62**, 1291 (2000); M. Nohara, M. Isshiki, H. Takagi, and R.J. Cava, *J. Phys. Soc. Jpn.* **66**, 1888 (1997); M. Nohara, H. Suzuki, N. Mangkorntong, and H. Takagi, *Physica* **C341–348**, 2177 (2000).
6. P. Pairor and M.B. Walker, *Phys. Rev.* **B65**, 064507 (2002); K. Maki, P. Thalmeier, and H. Won, *Phys. Rev.* **B65**, 140502R (2002); Hyun C. Lee and Han-Yong Choi, *Phys. Rev.* **B65**, 174530 (2002).
7. M.H.S. Amin, A.N. Omelyanchouk, S.N. Rashkeev, M. Coury, and A.M. Zagoskin, *Physica* **B318**, 162 (2002); Yu.A. Kolesnichenko, A.N. Omelyanchouk, and S.N. Shevchenko, *Fiz. Nizk. Temp.* **30**, 288 (2004) [*Low Temp. Phys*. **30**, 213 (2004)].